\documentclass[11pt,a4paper]{article}

\usepackage{jinstpub}

\usepackage[utf8]{inputenc} 
\usepackage[T1]{fontenc}    
\usepackage{hyperref}       
\usepackage{url}            
\usepackage{booktabs}       
\usepackage{amsfonts}       
\usepackage{nicefrac}       
\usepackage{microtype}      
\usepackage{lipsum}
\usepackage{graphicx}
\RequirePackage{doi}
\graphicspath{ {./images/} }
\newcommand{\pt}{p$_{\mathrm{T}}$}
\title{Jet Flavour Classification Using DeepJet}

\author[a]{Emil Bols, }
\author[b]{Jan Kieseler, }
\author[b]{Mauro Verzetti, }
\author[c,d]{Markus Stoye}
\author[e]{and Anna Stakia}

\affiliation[a]{Vrije Universiteit Brussels,\\Pleinlaan 2, 1050 Brussels, Belgium}
\affiliation[b]{CERN,\\ Espl. des Particules 1, 1211 Meyrin, Switzerland}
\affiliation[c]{Imperial College London,\\South Kensington Campus, London SW7 2AZ, UK}
\affiliation[d]{Reexen Technology}
\affiliation[e]{National Centre for Scientific Research 'Demokritos',\\ Patr. Gregoriou E \& 27 Neapoleos Str, Athens, Greece}
\emailAdd{emil.bols@cern.ch}

\arxivnumber{2008.10519}

\abstract{Jet flavour classification is of paramount importance for a broad range of applications in modern-day high-energy-physics experiments, particularly at the LHC. In this paper we propose a novel architecture for this task that exploits modern deep learning techniques. This new model, called DeepJet, overcomes the limitations in input size that affected previous approaches. As a result, the heavy flavour classification performance improves, and the model is extended to also perform quark-gluon tagging.}


\begin{document}
\maketitle
\flushbottom

\section{Introduction}
The Standard Model of particle physics (SM) \cite{Weinberg:1967tq, Salam:1968rm} is a remarkably effective theory, able to describe the experimental observations made thus far in high energy physics with unprecedented precision and completeness. Despite its success however, this model fails to explain several observations like the baryon asymmetry and the presence of dark matter, which inspires searches for extensions to the SM.
The study of the recently discovered \cite{Chatrchyan:2012xdj,Chatrchyan:2013lba,Aad:2012tfa} Higgs boson  \cite{Englert:1964et,Higgs:1964ia,Higgs:1964pj,Guralnik:1964eu, Higgs:1966ev,Kibble:1967sv}, and the search for extensions of the electroweak sector are two of the most active research sectors in the field. 

Because of the flavour asymmetry associated to production and decay processes in each case, the ability to classify jets originating from heavy-flavour (bottom and charm) quarks is important. Heavy-flavour (HF) jets contain an open-bottom or open-charm hadron as a result of the fragmentation process. This hadron carries a large fraction of the initial parton momentum. HF hadrons also have a sizeable lifetime, with a $c\tau$ of $\sim 0.5$ mm and $\sim 0.3$ mm for bottom and charm, respectively. Most of the HF hadrons produced in the fragmentation process undergo a decay process far enough from the primary interaction vertex (PV) to result in displaced tracks and their decay products can be clustered in resolved secondary vertices (SV).

Traditionally, jet flavour classification has leveraged both fragmentation and lifetime properties of HF hadrons. The discrimination power provided by single features is not adequate for a both efficient and pure classification, but the collective behaviour of the tracks in the jet, as well as the presence of a reconstructed secondary vertex, allows for a good performance of a combined algorithm.
For this reason, machine learning has long been used in the field of jet flavour classification, starting from simple naive bayes classifiers~\cite{CMS:2009gxa,Chatrchyan:2012jua}, to more complex and modern models like shallow neural networks~\cite{CMS-PAS-BTV-15-001,ATLASBTag2016}, or tree ensembles~\cite{CMS-PAS-BTV-15-001}.

More recently, simple dense deep neural networks~\cite{Sirunyan:2017ezt} have been successfully employed for this classification task alongside recurrent neural networks (LSTM)~\cite{LSTM}, outperforming previous classifiers. The ATLAS \textit{RNN} algorithm~\cite{ATL-PHYS-PUB-2017-003}, which is based on a recurrent neural network, processes as a sequence a small number of selected high purity charged particle tracks associated with the jet, sorted by their impact parameter. It exceeds the performance of the IP3D algorithm~\cite{ATLASBTag2016}, which considers the same inputs, but does not fully take into account the correlations between the tracks. 

Within CMS, the DNN based b-tagger DeepCSV~\cite{CMSbtv2017} has been used as the baseline reference for b-tagging performance for quite a while. DeepCSV is a fully connected neural network consisting of 5 layers with 100 nodes each using high level features of selected tracks and vertices as input.  In contrast with the ATLAS RNN tagger, DeepCSV combines properties from selected tracks and secondary vertices directly, but does not employ sequence processing.

The common feature between these two approaches, however, is that both models use only a small subset of the charged jet constituents that pass stringent quality criteria in order to provide the cleanest and simplest environment possible for the classifier to perform its task.

A stringent selection comes with information loss and therefore potential performance degradation. It has recently been shown that architectures compressing and converting individual track features to a latent space can overcome these limitations. The ATLAS Deep Sets b-tagging algorithm~\cite{ATLAS:2020jip} relying on the idea of Deep Sets~\cite{deepsets}, where the features of the input tracks are transformed into a latent space representation before being combined by a simple per-feature sum, has been shown to have similar performance as the RNN approach with reduced training and inference time given the same inputs. Studies on relaxing the input selection criteria on the tracks show a significant gain.

In this paper, we describe DeepJet, a new network architecture that does not rely on a selection of the jet constituents and thus overcomes the previous limitations on the purity and number of inputs. In addition, DeepJet considers the full information of all jet constituents, charged and neutral particles, secondary vertices, and global event variables simultaneously.

\section{Setup}
\label{sec:headings}
\subsection{Training Samples}

The DeepJet model is trained and tested using a sample of simulated anti k$_T$~\cite{Cacciari:2008gp} jets (with $R = 0.4$) drawn from a mix of QCD multijet events and fully hadronic top-quark pair ($t\bar{t}$) events, generated with PYTHIA8~\cite{Sjostrand:2014zea} and POWHEGv2~\cite{Nason:2004rx, Alioli:2010xd, Frixione:2007vw, Campbell:2014kua}, respectively. In both samples the hadronization and showering are performed using PYTHIA8. 
In order to apply the approach to a realistic reconstruction problem, GEANT4~\cite{Agostinelli:2002hh} is used to simulate the response of the CMS Phase 1 detector \cite{Chatrchyan:2008zzk, CMS:2012sda} to the generated particles. The jet constituents are reconstructed using the Particle Flow (PF) algorithm~\cite{Sirunyan:2017ulk} within the CMS Phase 1 detector reconstruction algorithms~\cite{cmssw_github}.
The entire training sample consists of roughly 130 million jets in total. This is then split into a training, a validation and a testing sample in the ratio 0.765:0.135:0.1.

\subsection{Labeling}
The jets are labelled by \textit{ghost association}~\cite{CMSbtv2017}. In this approach the last instance of each generated b and c hadron before the decay have their momentum scaled to a very small value, in order to carry only directional information. These "ghosts" are then included among the particles to be clustered by the jet algorithm. If a jet contains at least one b hadron, it is labeled a b jet. If the jet contains at least one c hadron, and no b hadrons, it is labelled a c jet. Everything else is then labelled as a light jet. The jets labelled as b jets are further sub-labelled into three categories: $bb$ for jets containing two b hadrons, $b_{lept}$ for jets containing b hadrons decaying leptonically, and $b$ for jets with b hadrons decaying hadronically. The light jets are also sub-labelled into quark and gluon jets using the CMS definition~\cite{CMS-PAS-JME-16-003}.

\subsection{Input features}

DeepJet uses approximately 650 input variables, divided into four categories: global variables, charged PF candidate features, neutral PF candidate features, and SV features associated with the jet. The global variables include jet-level information such as the jet kinematics, the number of tracks in the jet, as well as the number of secondary vertices in the jets. 
Additionally, the global variables also include the number of reconstructed primary vertices in the event, in order to help the network learn the effect of multiple interactions per bunch crossing ({pileup}). 
The information related to the charged PF candidates is summarised in 16 features of the first 25 candidates with respect to their displacement significance. These features contain information on the track kinematics, track fit quality, displacement, and displacement uncertainty with respect to the PV.
Similarly, 6 features of the leading 25 neutral candidates are provided to the network.
Finally, up to 4 secondary vertices are considered, each with 12 variables, such as the flight distance significance. A full list of the variables used can be found in the appendix. 

\subsection{Preprocessing of features}

Since the b-tagger is meant to generalise beyond the specific kinematic domain used for the training, 
it is important to remove any flavour dependence on transverse momentum (\pt), or pseudorapidity ($\eta$) to avoid the classifier leveraging those differences.
During the data pre-processing, we downsample each of the flavour classes in \pt\ and $\eta$ such that they have the same shape as the $b$ jet class. Several variables are also bounded within a physically well defined range. In case features of an object are either not available or infinite, they are replaced with an appropriately chosen value for the specific feature. 
The particle objects are ordered based on their assumed importance. Charged particle candidates are ordered by impact parameter significance, while neutral candidates are ordered by the shortest angular distance to a secondary vertex. If there is no secondary vertex in the jet, \pt\ ordering is used instead. Secondary vertices are sorted by flight distance significance.

\section{DeepJet}
\subsection{Architecture}

\begin{figure}
  \centering
  \includegraphics[width=1\textwidth]{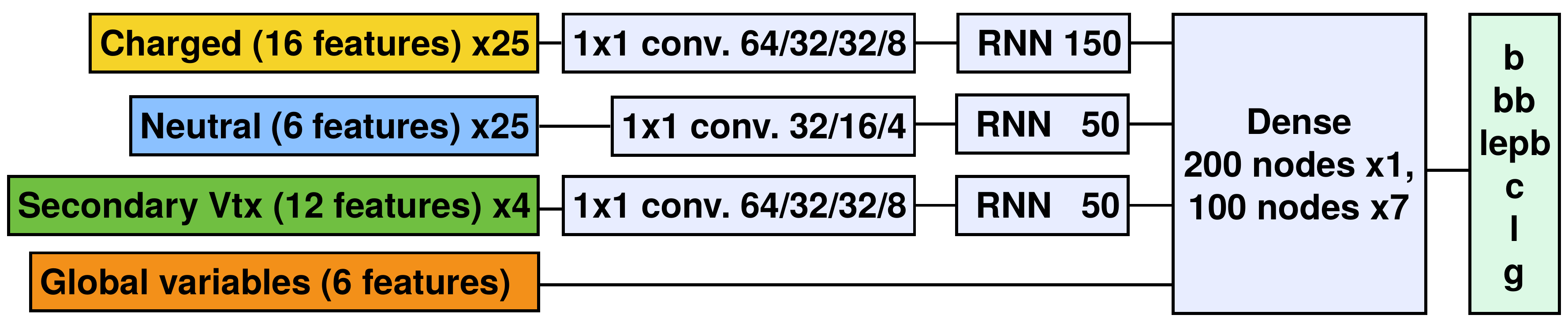}
  \caption{An illustration of the DeepJet architecture. Three seperate branches are used to process charged candidates, neutral candidates and secondary vertices. The algorithm makes use of 1x1 convolutional layers to perform automatic feature engineering for each class of jet constituents. The three RNN (LSTM) layers combine the information for each sequence of constituents. Finally the full jet information is combined using fully connected layers.  }
  \label{fig:ARCH}
\end{figure}

The concept of DeepJet is to use low-level features from as many jet constituents as possible as opposed to selecting a few that are well identified and reconstructed and exploiting higher level features. The exact number of features and constituents can vary with experiment and use case, but the all-inclusive concept remains the same. To process this large input space, a suitable architecture is needed. 
An illustration of the DeepJet architecture can be seen in Fig.~\ref{fig:ARCH}. In the first step, an automatic feature engineering is performed for each constituent using convolutional branches comprising 1x1 convolutional layers~\cite{lecun-01a}. The filter size of 1x1 is chosen as inherently all candidates should undergo the same feature transformation without taking into account the other constituents in the jets, and a priori a specific class of physics objects should be treated in a consistent manner independent of input ordering.\footnote{This concept is closely related to the idea of Deep Sets~\cite{deepsets}.}
Separate convolutional branches are used for vertices, charged PF candidates and neutral PF candidates. The convolutional branches use a sequences of layers with a decreasing number of filters, as this projects and compresses the features of each constituent into a lower dimensional space. Each of the outputs are then fed into a a dedicated recurrent layer of the LSTM type. This treats the constituents as a sequence, following the order described in section 2.4.
In particular the LSTM nodes are well suited for dealing with the variable length sequences of inputs, which occurs in jet physics. The charged candidate LSTM uses 150 nodes, whereas the neutral LSTM and the secondary vertex LSTM each use 50 nodes. The three LSTM outputs are concatenated with the global features of the jet and then fed into a fully connected layer with 200 nodes, followed by 7 fully connected layers with 100 nodes each. Finally 6 outputs nodes are integrated into the multi-classifier, allowing it to perform b-tagging, c-tagging and quark/gluon tagging.  Throughout the network ReLU \cite{ReLU} activation functions are used, except for the output layer, where the softmax activation function is applied. 

At the beginning of the network, and in between each layer, 

batch normalization \cite{DBLP:journals/corr/IoffeS15} is performed, and dropout~\cite{dropout} with a rate of 0.1 is applied for regularisation.

\subsection{Training procedure}

The neural network is trained using the Adam optimizer~\cite{kingma2014adam} with a learning rate of $3\cdot 10^{-4}$ for 65 epochs and a categorical cross entropy loss. The learning rate is also halved if the validation sample loss stagnates for more than 10 epochs.
After the first training epoch, the input batch normalization layer is frozen. Possible over-training is monitored through the validation loss, which is mostly dominated by medium \pt\ jets, and additionally through ROC curves in different \pt\ ranges. 
The training is performed on an Nvidia GEFORCE GTX 1080 Ti GPU. With a training sample size of roughly 130 million examples, the training convergence time is around 3 days.  

\section{Performance}

\subsection{Comparison with other b-tagging algorithms}

\begin{figure}
  \centering
  \includegraphics[width=0.45\textwidth]{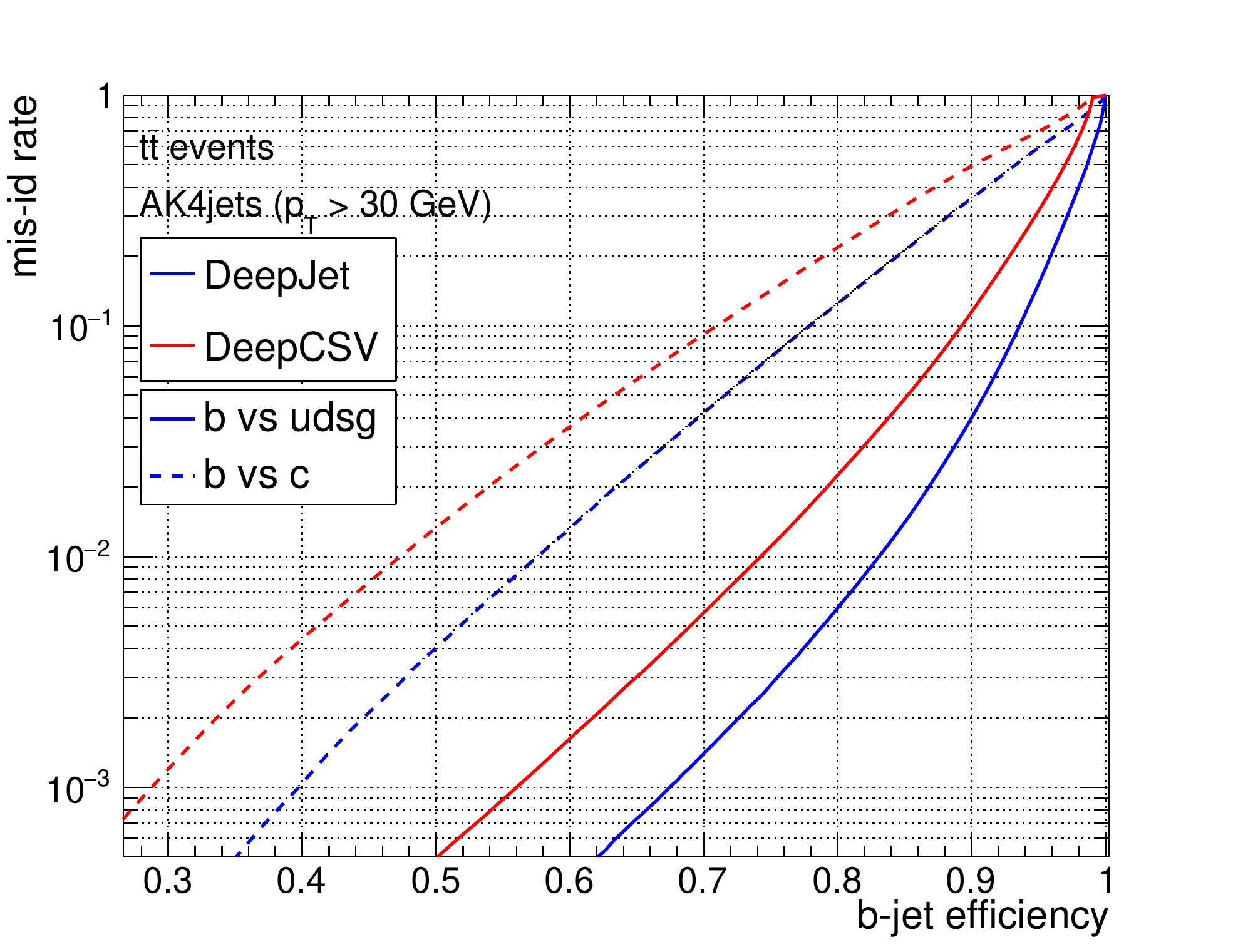}
  \includegraphics[width=0.45\textwidth]{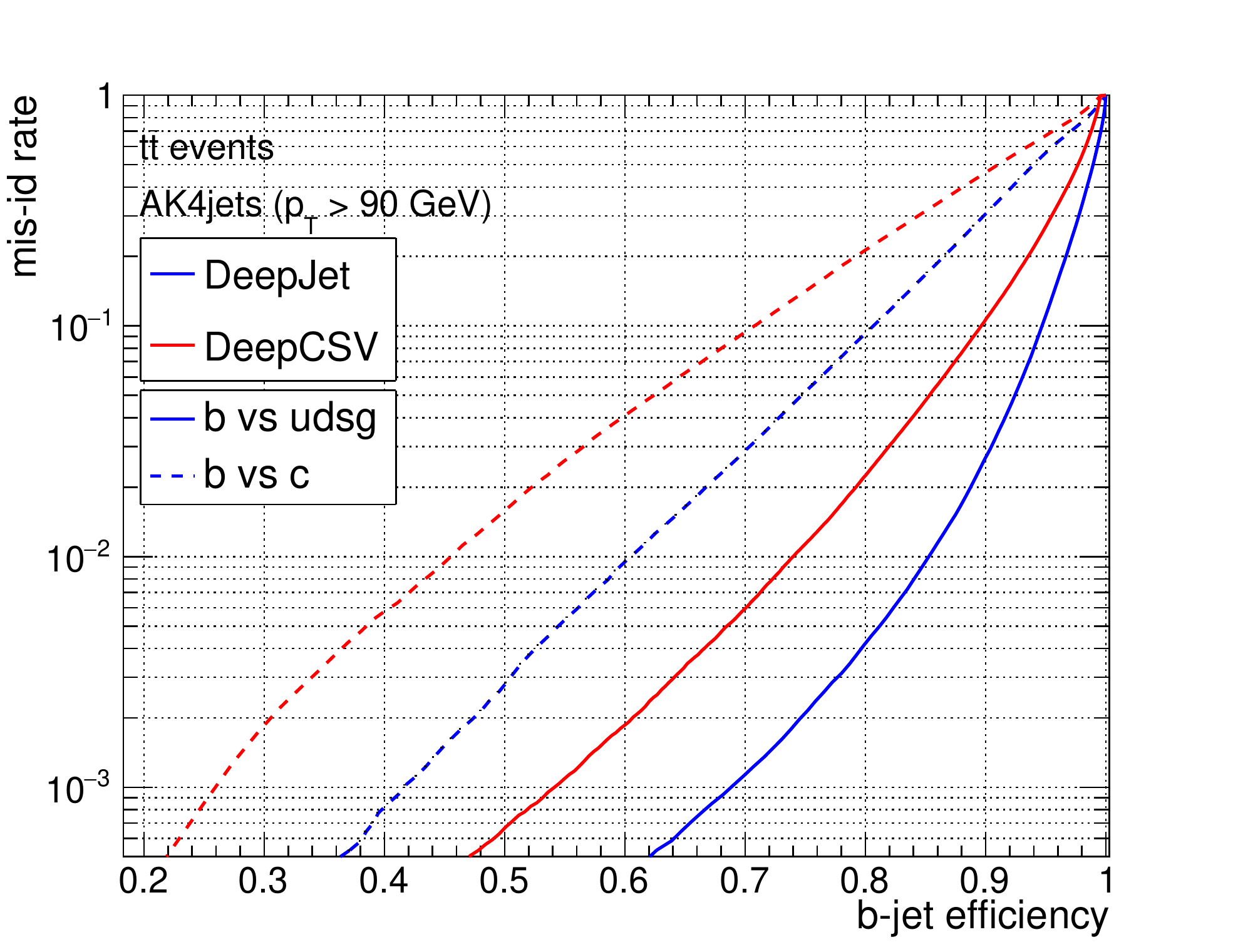}
  \caption{
  Performance of the DeepJet and DeepCSV b-tagging algorithms on $t\bar{t}$ events with both top quarks decaying hadronically. The jets are required to have \pt\ $> 30$ GeV (left) and \pt\ $> 90$ GeV (right). The performance is shown for both b vs. c classification (dashed lines), and b vs. light (solid lines).
  }
  \label{fig:TT_eff_pt_ROCS}
\end{figure}

The overall performance of DeepJet is compared to the CMS b-tagger DeepCSV, which is a multi-classifier embedded in the CMS reconstruction framework, and which uses similar, but strongly preselected, inputs along with additional high-level variables.
The comparison is made on a fully hadronic $t\bar{t}$ sample, as shown in Fig.~\ref{fig:TT_eff_pt_ROCS}. DeepJet performs significantly better than DeepCSV with an efficiency\footnote{True Positive Rate.} increase of almost 20$\%$ at $10^{-3}$ misidentification probability\footnote{False Positive Rate.} for jets with \pt\ $>90\mathrm{GeV}$.

\begin{figure}[htpb]
  \centering
  \includegraphics[width=0.45\textwidth]{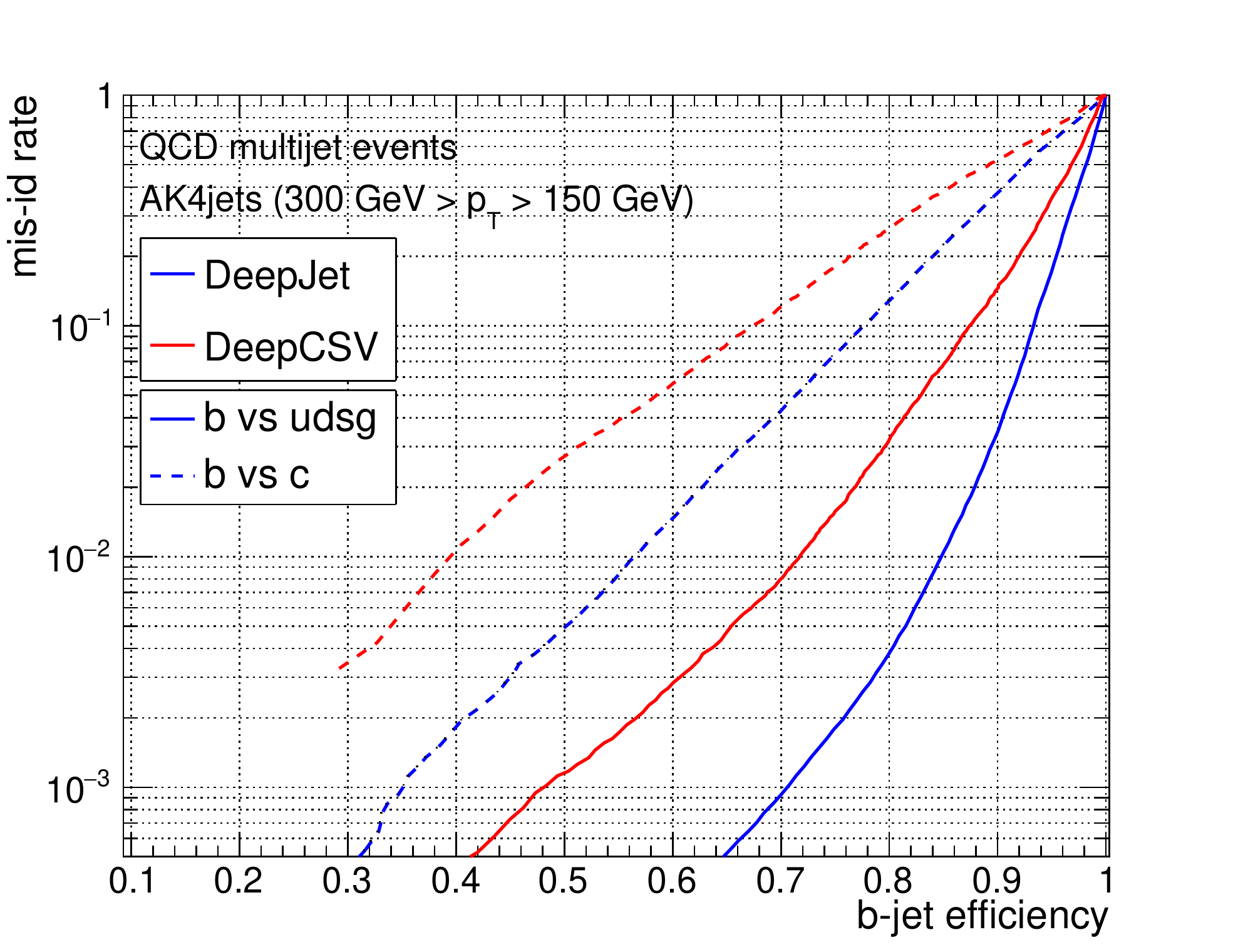}
  \includegraphics[width=0.45\textwidth]{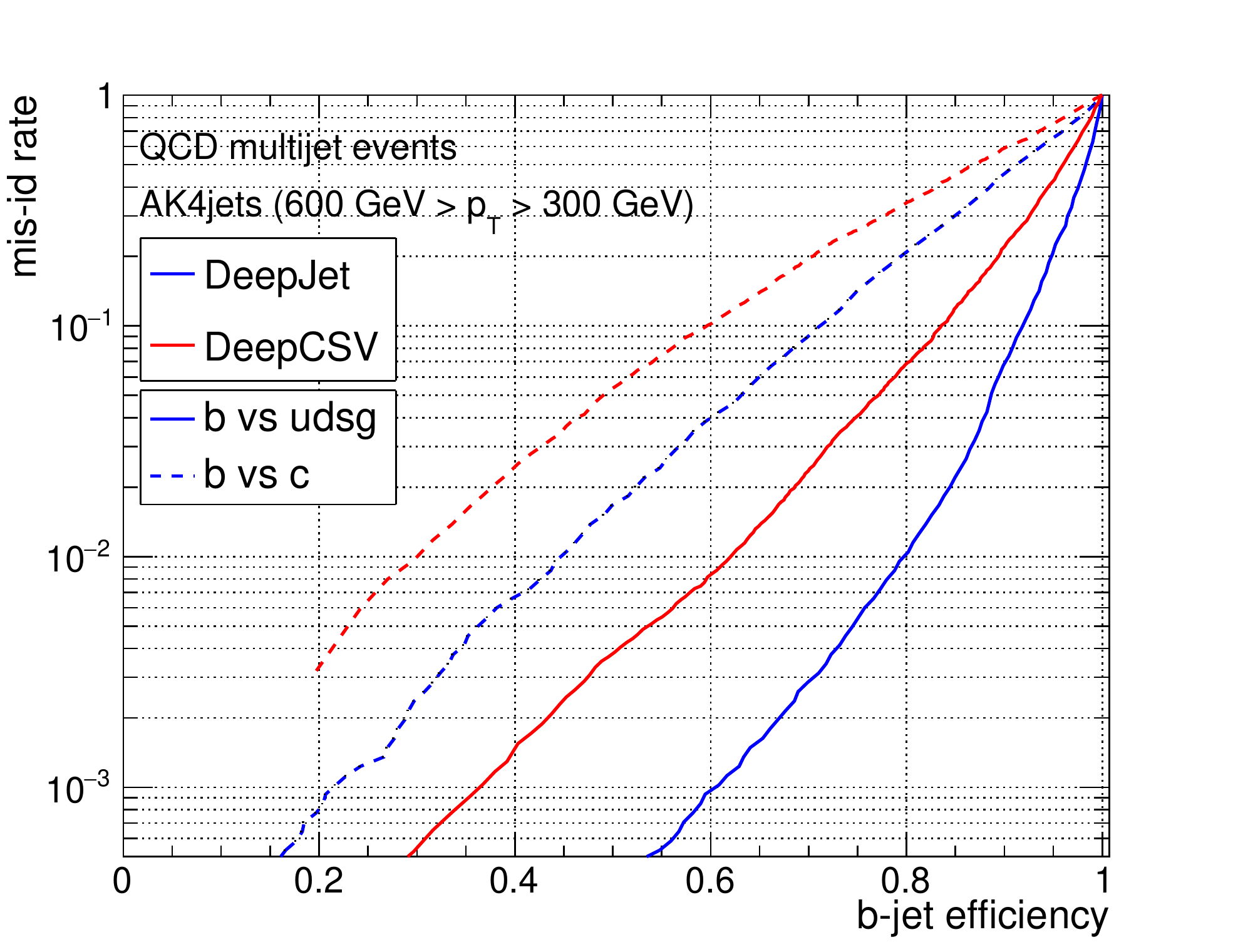}
  \includegraphics[width=0.45\textwidth]{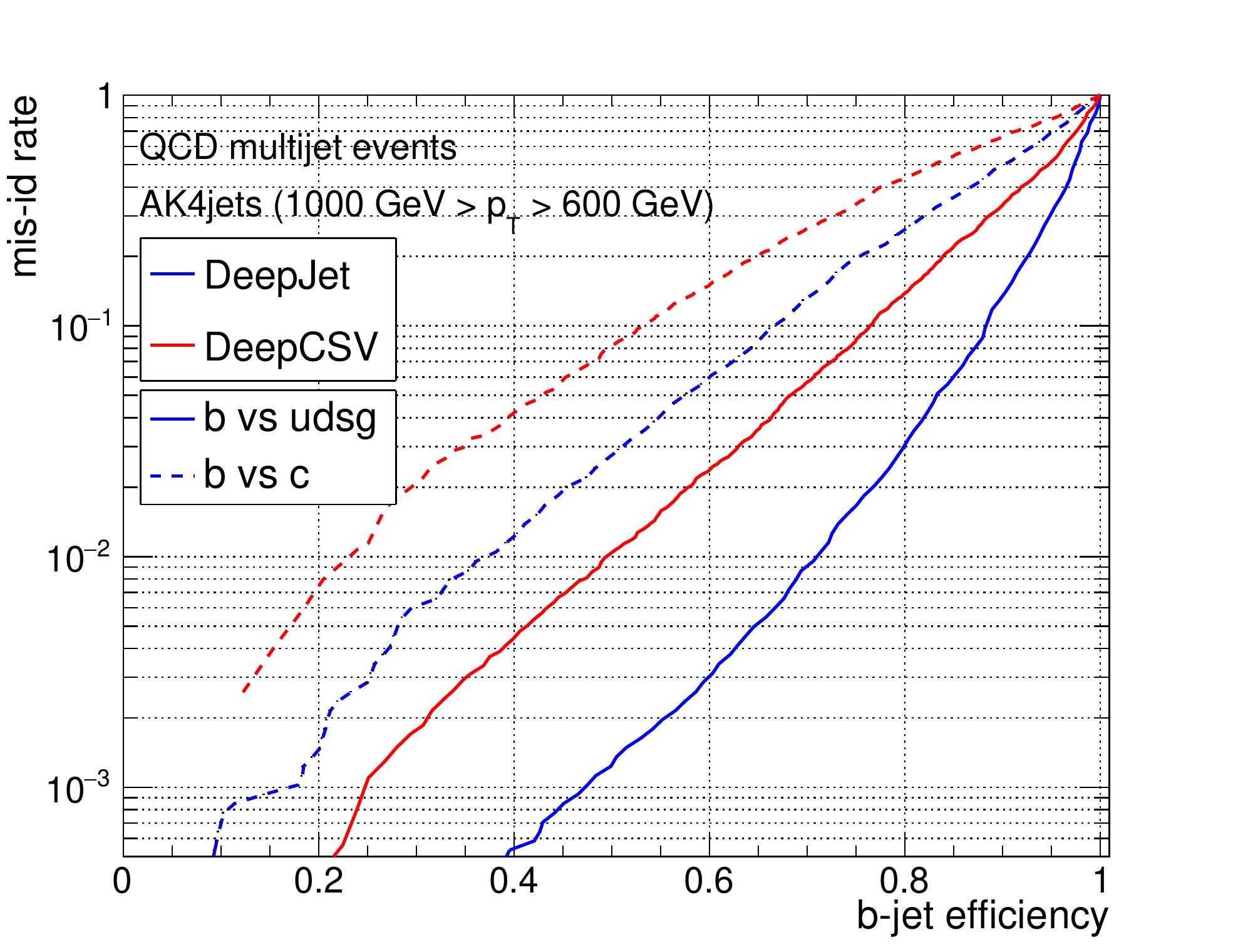}
  \caption{The performance of DeepJet and DeepCSV in three different jet \pt\ ranges in QCD multijet events. The performance is shown for both b vs. c (dashed lines), and b vs. light (solid lines).}
  \label{fig:QCD_eff_pt_ROCS}
\end{figure}

\begin{figure}[htpb]
  \centering
  \includegraphics[width=0.7\textwidth]{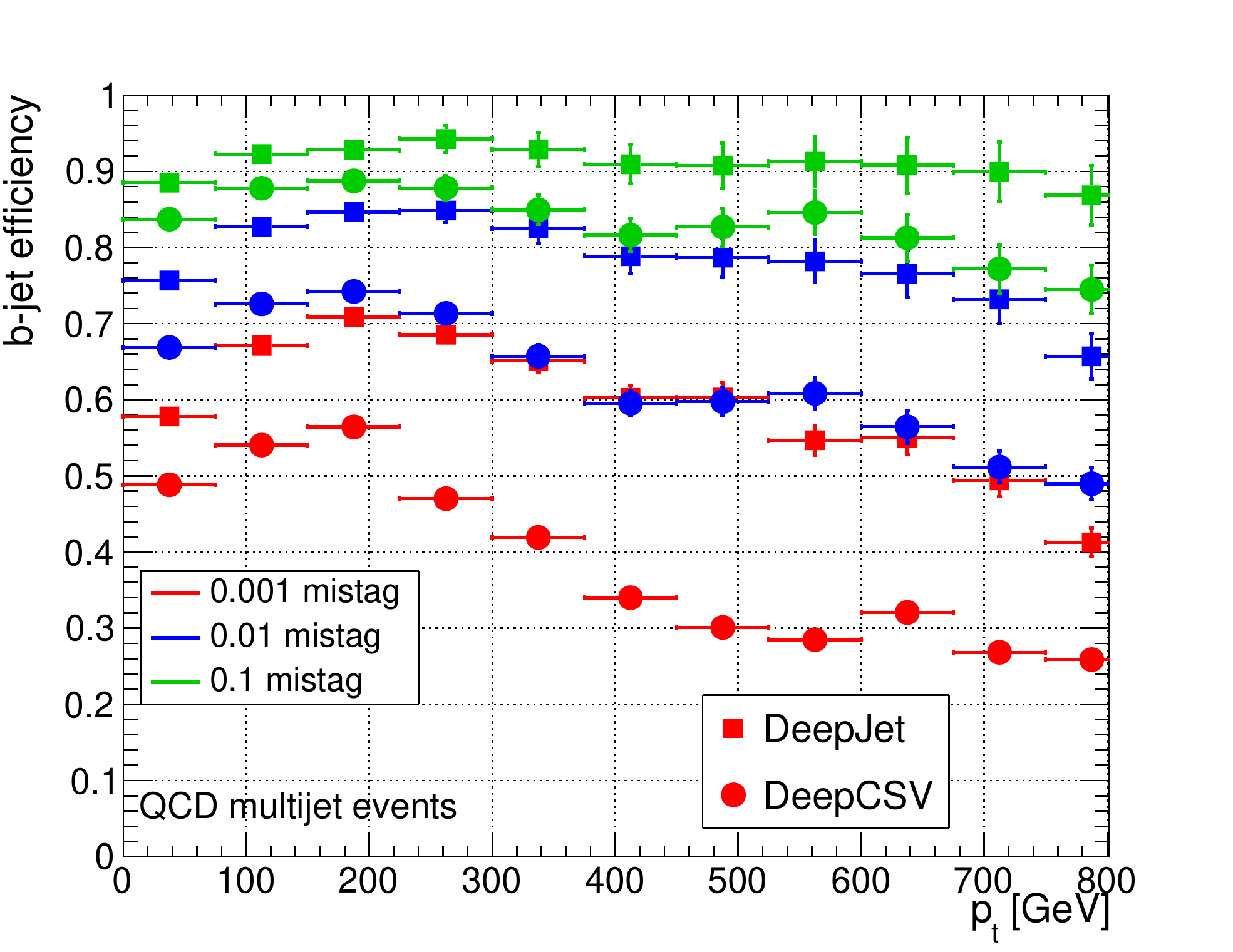}
  \caption{Performance of DeepJet and DeepCSV as a function of jet \pt. The efficiency for b jets is shown for three different fixed light jet misidentification probabilities (mistag rates) using QCD multijet events.}
  \label{fig:QCD_eff_pt}
\end{figure}

The performance of the classifiers can also be compared on a set of multi-jet QCD events, which allow accessing higher jet momenta. The inclusive results are shown in Fig.~\ref{fig:QCD_eff_pt_ROCS}. A comparison for different jet \pt\ can be found in Fig.~\ref{fig:QCD_eff_pt} for fixed light jet misidentification probabilities. In both cases DeepJet shows an increasing performance gain, in particular for higher jet \pt, presumably exploiting the information contained in all constituents.

\begin{figure}[htpb]
  \centering
  \includegraphics[width=0.7\textwidth]{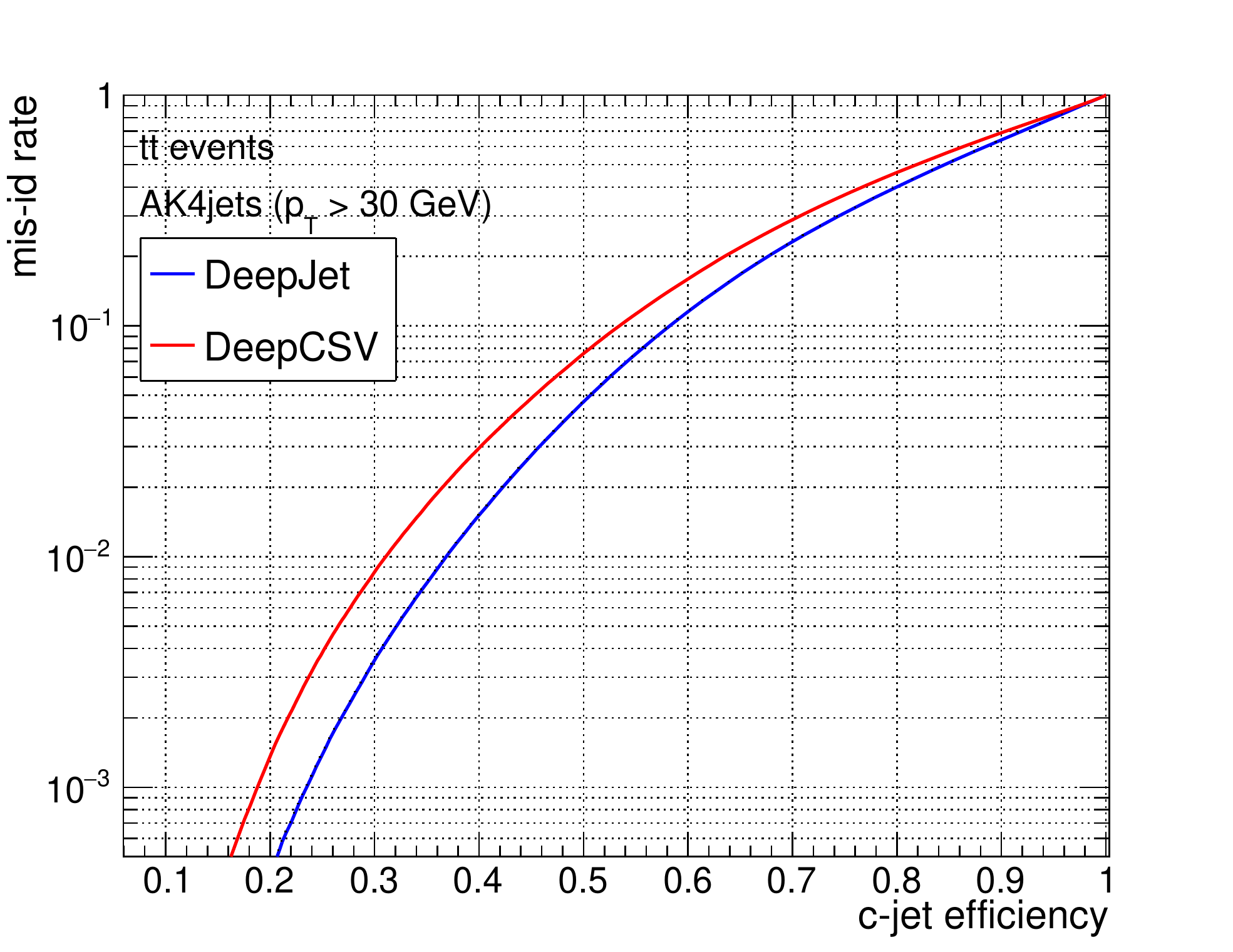}
  \caption{Light vs. c jet classification performance of DeepJet and DeepCSV in fully hadronic $t\bar{t}$ events.}
  \label{fig:ctag_ROCS}
\end{figure}

The multi-classifier nature of DeepJet and DeepCSV allows the models to also efficiently perform c jet identification. Figure~\ref{fig:ctag_ROCS} shows that DeepJet outperforms DeepCSV also in this task. In this case, both DeepCSV and DeepJet use a binary discriminator defined as $\frac{P(\mathrm{c})}{P(\mathrm{c})+P(\mathrm{uds})+P(\mathrm{g})}$, where $P(\mathrm{c})$, $P(\mathrm{uds})$, and $P(\mathrm{g})$ describe the classifier output for c, light quark (u, d and s quark), and gluon jets, respectively.

\begin{figure}[htpb]
  \centering
  \includegraphics[width=0.7\textwidth]{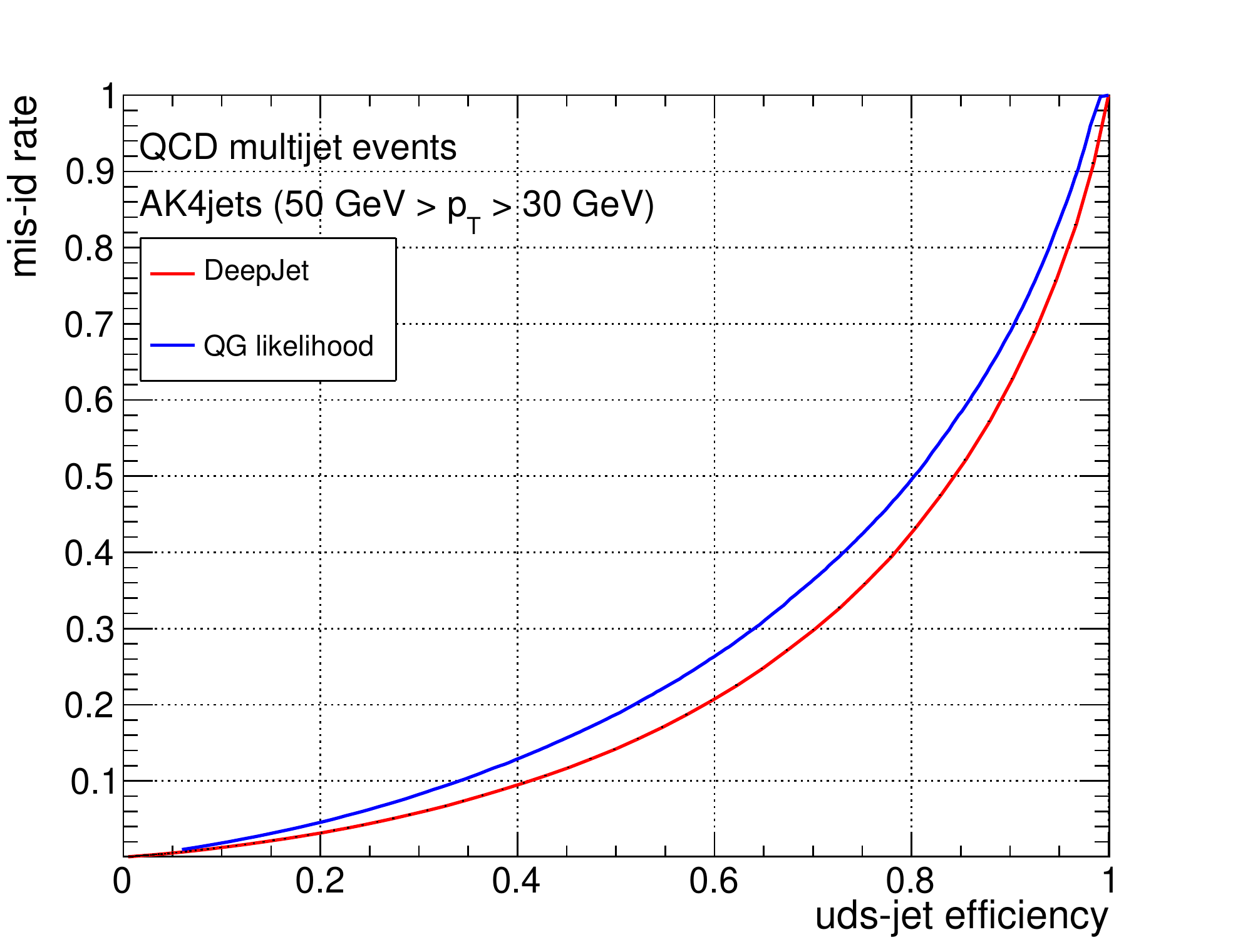}
  \caption{Quark gluon discrimination performance of DeepJet compared to the CMS ``quark-gluon likelihood'' method in a sample of pure light quark (uds) and gluon jets.}
  \label{fig:qg_ROCS}
\end{figure}

Finally, DeepJet can be used for quark/gluon discrimination. DeepJet is compared to the ``quark/gluon likelihood''~\cite{CMS-PAS-JME-16-003}, a binary quark/gluon classifier included in the CMS reconstruction framework. For this purpose, the DeepJet output is also projected to a binary classifier, $\frac{P(\mathrm{uds})}{P(\mathrm{uds})+P(\mathrm{g})}$. 
Also here, a performance increase can be observed when employing the DeepJet model on a sample consisting purely of light quark and gluon jets, as shown in Fig.~\ref{fig:qg_ROCS}. Moreover, given that a standard quark/gluon classifier uses a binary approach, we expect a larger performance gain in scenarios where gluon jets and light quark jets also need to be separated from heavy flavour jets.

\begin{figure}[htpb]
  \centering
  \includegraphics[width=0.7\textwidth]{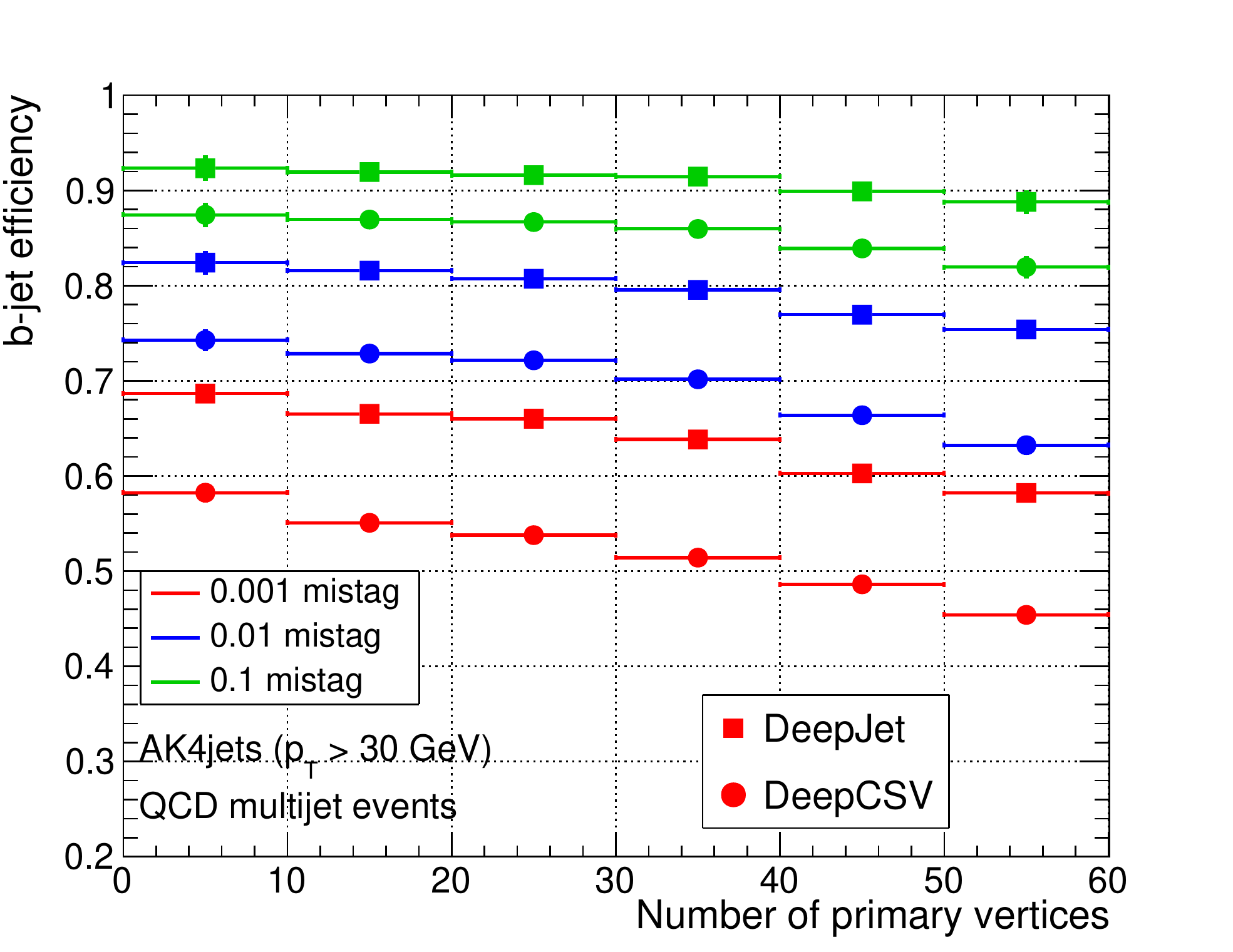}
  \caption{Performance of DeepJet and DeepCSV as a function of reconstructed primary vertices. The efficiency for b jets is shown for three different fixed light jet misidentification probabilities (mistag rates) using QCD multijet events.}
  \label{fig:pileup}
\end{figure}

The effects of multiple interactions per bunch crossing ({pileup}) can further degrade the b-tagging performance. In Fig.~\ref{fig:pileup} the b-tagging performance is shown as a function of number of reconstructed primary vertices for both DeepJet and DeepCSV. Despite the use of more low purity input in DeepJet, a similar level of performance degradation is observed in both algorithms when increasing pileup.

\subsection{Qualitative assessment of the performance gain source}

\begin{figure}[htpb]
  \centering
  \includegraphics[width=0.85\textwidth]{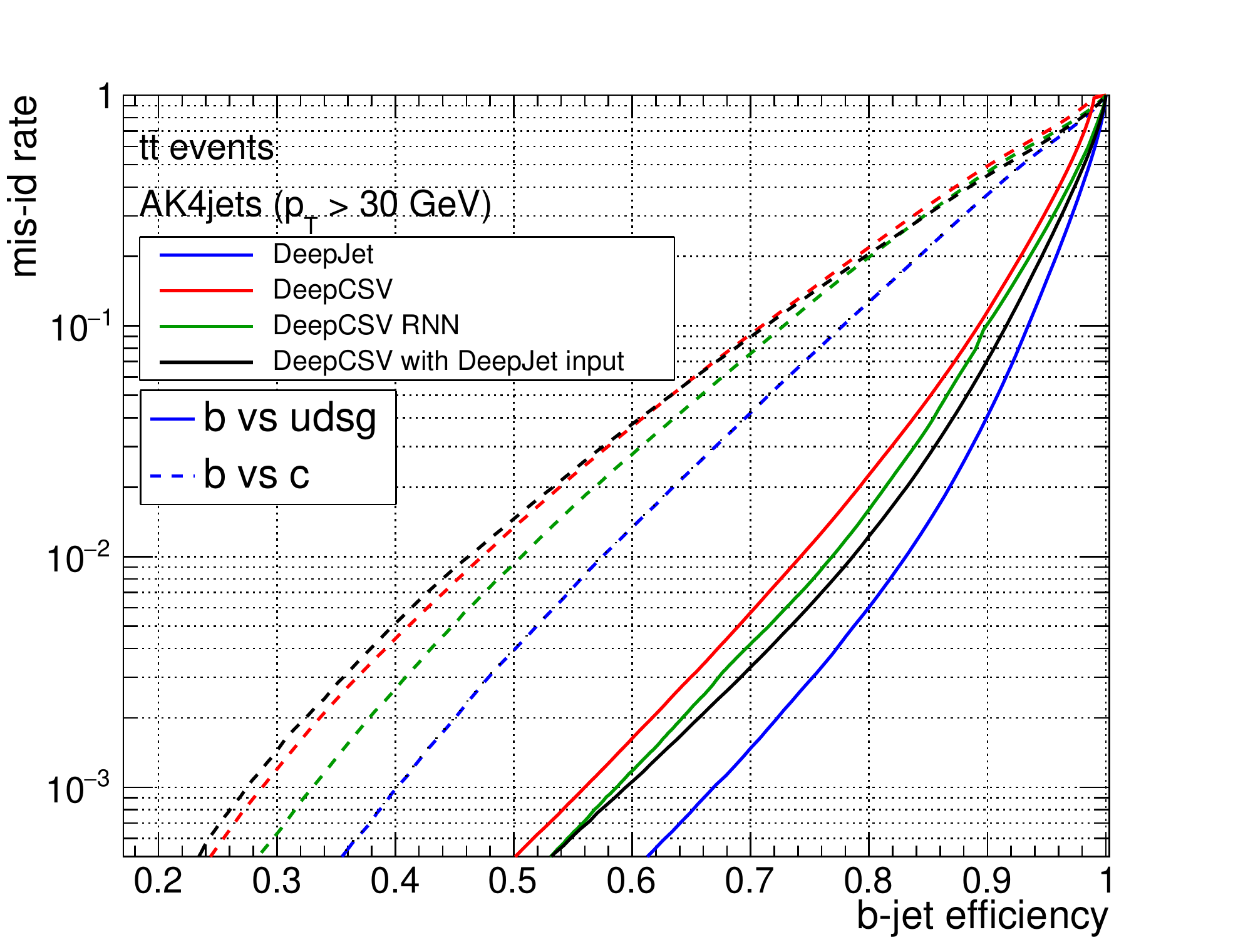}
  \caption{An interpolation between the DeepCSV tagger \cite{CMSbtv2017} and DeepJet.
  }
  \label{fig:DeepCSVtoDeepJet}
\end{figure}

Given that the architecture and the network inputs are changed simultaneously between DeepCSV and DeepJet, two new network configurations are trained to assess the relative impact of these changes to the performance. The first model, referred to as DeepCSV RNN, contains the same inputs as DeepCSV, but its architecture is changed to mimic that of DeepJet, with a series of convolutional layers feeding into a LSTM processing the list of inputs belonging to different tracks. The second model uses the architecture of DeepCSV but uses the full input of DeepJet. The performance of these models are compared with DeepCSV and DeepJet in Fig.~\ref{fig:DeepCSVtoDeepJet} for inclusive $t\bar{t}$ events. Increasing the number of inputs does not help much, when it is not complemented by a network structure that allows efficient processing of the larger and less pure track set. However adopting the new architecture without increasing the input and removing the track selection also has limited gain. Similar improvements coming with similar preprocessing of unselected jet constituent inputs could be observed in the Deep Set based tagger~\cite{ATLAS:2020jip}.
This qualitative statement also explains the additional gain observed in DeepJet at high jet \pt\, as the track selection was historically optimised for mild jet boosts.

The default DeepJet architecture is also compared to an adapted architecture employing the Deep Set idea. In the DeepSet approach it is required to project the single objects into a high dimensional feature space, whereas the DeepJet model compresses the features.
To mimic DeepSet, the DeepJet architecture is adapted as follows: the 1x1 convolutional branches are extended to contain 100 filters, each, increasing the number of free parameters. The final layer of each branch is enlarged further, with the charged branch being set to 256 filters, whereas the vertex and the neutral branch are set to 128 filters. At the same time, the LSTM layers are replaced by a simple sum that is evaluated independently for each convolutional branch, after which these sums are concatenated and fed to a series of fully connected layers. Another comparison model is added that shows a DeepJet model with the same enlarged convolutional structure as the DeepSet model, but keeping the LSTM for aggregating the tracks.

\begin{figure}[htpb]
  \centering
  \includegraphics[width=0.49\textwidth]{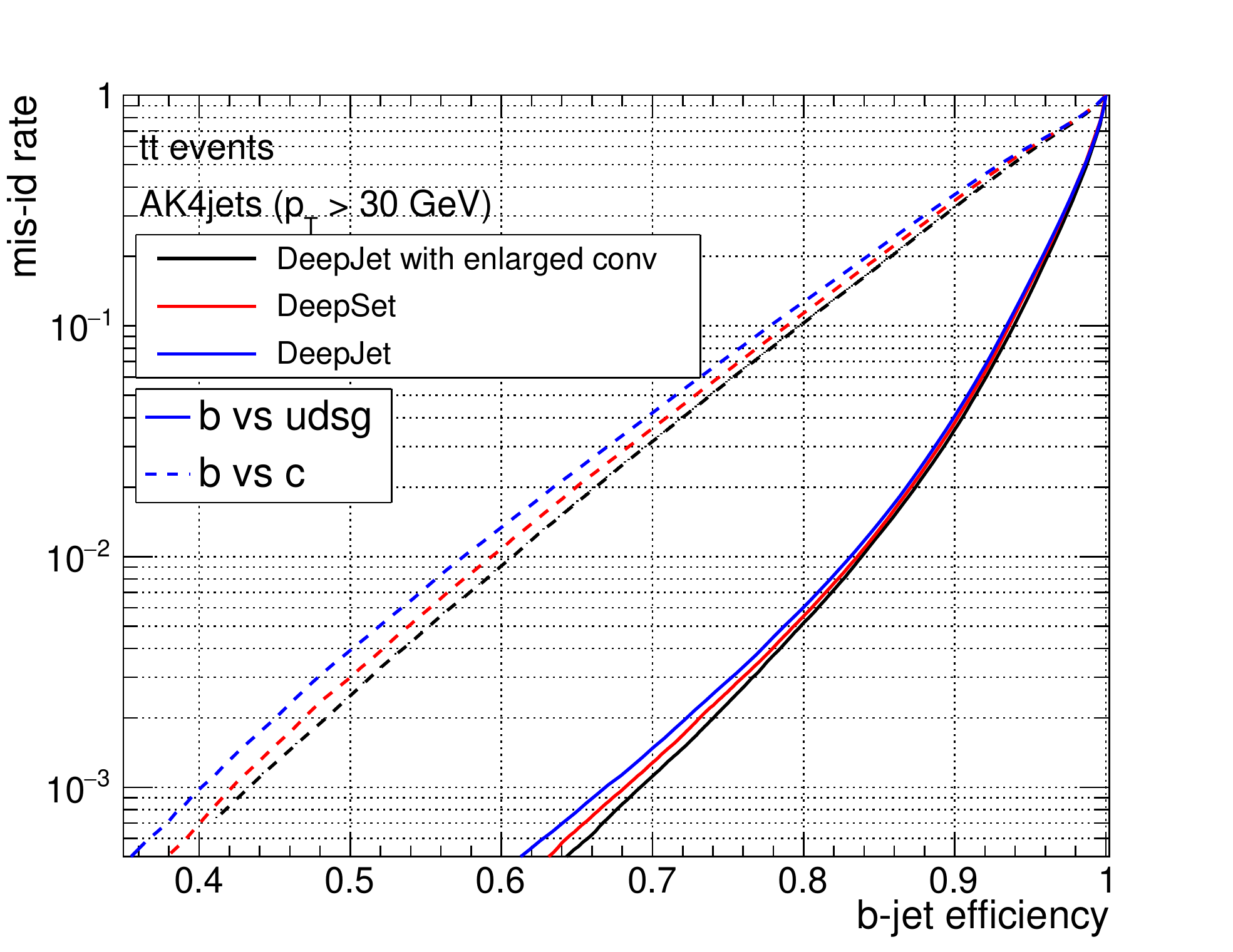}
  \includegraphics[width=0.49\textwidth]{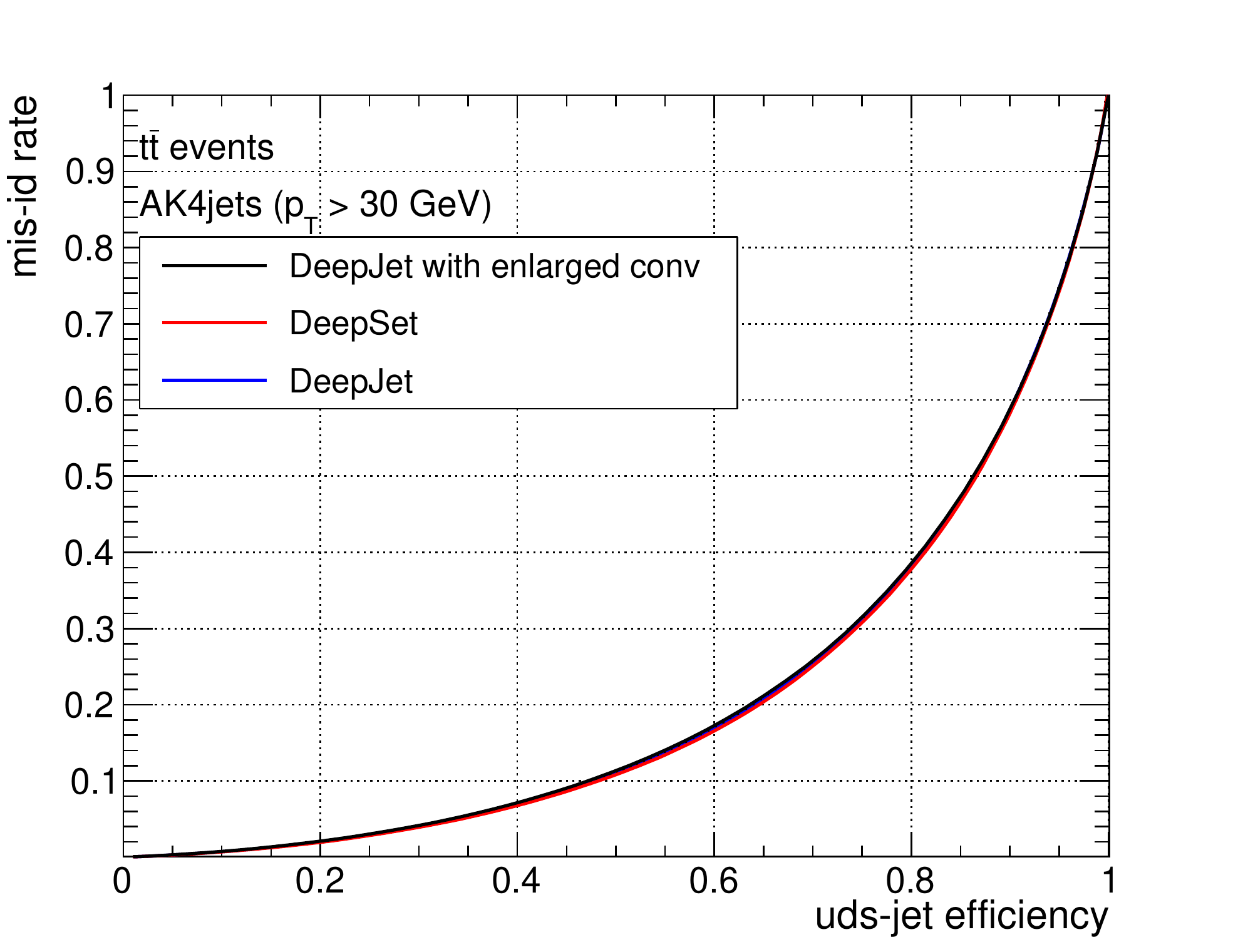}
  \caption{Comparison of the DeepJet architecture, an adapted DeepJet architecture based on the Deep Sets approach, and a DeepJet model with enlarged number of convolutional filters. Left: b jet identification performance. Right: quark/gluon jet separation performance.
  }
  \label{fig:deepset_ROCS}
\end{figure}

As shown in Fig.~\ref{fig:deepset_ROCS}, the performance of the Deep Set based architecture shows a slight gain for b to c jet identification, but shows no significant difference with respect to the b to light jet identification  or the quark/gluon discrimination performance compared to the default DeepJet architecture. The DeepJet model using enlarged convectional filters performs slightly better than DeepSet, but like DeepSet, it projects the input objects into large features spaces, leading to increased resource requirements for the subsequent LSTMs. 

\section{Conclusion}

A multiclass flavour tagging algorithm, DeepJet, has been developed to exploit the full information in a jet. The model was tested using CMS simulation. The model relies on low-level variables and loose selection of the inputs, and it uses an architecture that can process these inputs in an efficient manner. When compared with fully connected models using a smaller set of engineered features a gain in performance is observed in all topologies of flavour tagging, in some cases exceeding a two-fold efficiency gain for the same misidentification rate. The model goes beyond heavy flavour tagging and performs quark-gluon discrimination as well. The applicability of the DeepJet approach has been successfully extended in different directions, such as the construction of tagging algorithms for exotic long lived particles \cite{CMS:LLP}.

DeepJet has been trained using the homonymous github package \cite{deepJetGit, deepJetCoreGit}, with a streamlined training process and multi-threaded data access. The training set has been scraped from the simulation data using \cite{deepntuples}, which contains an algorithmic description of all the variables used in the training.

\section*{Acknowledgements}
We thank our colleagues within the CMS collaboration for the ability to use the CMS simulation, and especially the CMS b-tagging and vertexing community for their continuous support to integrate DeepJet in the CMS framework. The training of the model was performed on the GPU clusters of the CERN EP/CMG group.

\bibliographystyle{JHEP}
\bibliography{references}  




\newpage
\appendix
\section{Appendices: Inputs}
\subsection{List of global variables}
\begin{itemize}
\item Jet $p_t$
\item Jet $\eta$
\item The number of charged particle flow candidates in the jet
\item The number of neutral particle flow candidates in the jet
\item The number of secondary vertices in the jet
\item The number of primary vertices in the event
\end{itemize}

\subsection{List of charged candidate variables}
\begin{itemize}
\item Charged track $\eta$ relative to the jet axis
\item Charged track $p_t$ relative to the jet axis
\item Dot product of the jet and track momentum
\item Dot product of the jet and track momentum divided by the magnitude of the jet momentum
\item $\Delta$R between the jet axis and the track
\item The track 2D impact parameter value
\item The track 2D impact parameter significance
\item The track 3D impact parameter value
\item The track 3D impact parameter significance
\item The track distance to the jet axis
\item Fraction of the jet momentum carried by the track.
\item $\Delta$R between the track and the closest secondary vertex
\item An integer flag that indicate whether the track was used in the primary vertex fit.
\item The charged candidates PUPPI weight
\item $\chi^2$ of the charged track fit.
\item A integer flag which indicate the quality of the fitted track, based on number of detector hits used for the reconstruction as well as the overall $\chi^2$ of the charged track fit.

\end{itemize}

\subsection{List of neutral candidate variables}
\begin{itemize}
\item Fraction of the jet momentum carried by the neutral candidate
\item $\Delta$R between the jet axis and the neutral candidate
\item A integer flag indicating whether the neutral candidate is a photon.
\item Fraction of the neutral candidate energy deposited in the hadronic calorimeter.
\item $\Delta$R between the neutral candidate and the closest secondary vertex
\item The neutral candidates PUPPI weight
\end{itemize}

\subsection{List of secondary vertex variables}
\begin{itemize}
\item Secondary vertex $p_t$
\item $\Delta$R between the jet axis and the secondary vertex
\item Secondary vertex mass
\item Number of tracks in the secondary vertex
\item $\chi^2$ of the secondary vertex fit
\item Reduced $\chi^2$ of the secondary vertex fit
\item The secondary vertex 2D impact parameter value
\item The secondary vertex 2D impact parameter significance
\item The secondary vertex 3D impact parameter value
\item The secondary vertex 3D impact parameter significance
\item Cosine of the angle between the secondary vertex flight direction and the direction of the secondary vertex momentum.
\item Ratio of the secondary vertex energy to the jet energy
\end{itemize}

\end{document}